\documentclass[a4paper]{jpconf}
\usepackage{graphicx}
\newcommand{\solm}{M$_{\odot}$\ }

\begin{document}

\title{Coordinated multi-wavelength observations of Sgr A*}

\author{A. Eckart$^{1,2}$,
       R. Sch\"odel$^{3}$,
       F. K. Baganoff$^4$,
       M. Morris$^5$,
       T. Bertram$^1$,
       M. Dovciak$^{6}$, 
       D. Downes$^7$, 
       W.J. Duschl$^{8,9}$,
       V. Karas$^{6}$, 
       S. K\"onig$^{1,2}$,
       T. Krichbaum$^{2}$,
       M. Krips$^{10}$,
       D. Kunneriath$^{1,2}$,
       R.-S. Lu$^{2,1}$,
       S. Markoff$^{11}$,
       J. Mauerhan$^5$,
       L. Meyer$^5$,
       J. Moultaka$^{12}$,
       K. Muzic$^{1,2}$,
       F. Najarro$^{13}$,
       K. Schuster$^7$,
       L. Sjouwerman$^{14}$,
       C. Straubmeier$^1$,
       C. Thum$^7$,
       S. Vogel$^{15}$,
       H. Wiesemeyer$^{16}$, 
       G. Witzel$^1$,
       M. Zamaninasab$^{1,2}$,
       A. Zensus$^{2}$
}

\address{
$^1$  University of Cologne, Z\"ulpicher Str. 77, D-50937 Cologne, Germany\\
$^2$  Max-Planck-Institut f\"ur Radioastronomie, Auf dem H\"ugel 69, 
    53121 Bonn, Germany\\
$^3$  Instituto de Astrof\'isica de Andaluc\'ia, Camino Bajo de
    Hu\'etor 50, 18008 Granada, Spain \\
$^4$ Center for Space Research, Massachusetts Institute of
            Technology, Cambridge, MA~02139-4307, USA \\
$^5$  Department of Physics and Astronomy, University of California, 
     Los Angeles, CA 90095-1547, USA\\
$^6$ Astronomical Institute, Academy of Sciences, 
        Bo\v{c}n\'{i} II, CZ-14131 Prague, Czech Republic \\
$^7$ Institut de Radio Astronomie Millimetrique, Domaine Universitaire, 
    38406 St. Martin d'Heres, France\\
$^8$ Institut f\"ur Theoretische Physik und Astrophysik,
        Christian-Albrechts-Universit\"at zu Kiel, Leibnizstr. 15
        24118 Kiel, Germany \\
$^9$ Steward Observatory, The University of Arizona, 933 N. 
     Cherry Ave. Tucson, AZ 85721, USA\\
$^{10}$ Harvard-Smithsonian Center for Astrophysics, SMA project, 
     60 Garden Street, MS 78 Cambridge, MA 02138, USA\\
$^{11}$    Astronomical Institute `Anton Pannekoek', 
        University of Amsterdam, Kruislaan 403,
        1098SJ Amsterdam, the Netherlands\\
$^{12}$ Observatoire Midi-Pyr\'en\'ees,
        14, Avenue Edouard Belin, 31400 Toulouse, France\\
$^{13}$ Instituto de Estructura de la Materia, 
        Consejo Superior de Investigaciones Cientificas, 
        CSIC, Serrano 121, 28006 Madrid, Spain\\
$^{14}$ National Radio Astronomy Observatory,
       PO Box 0, Socorro, NM 87801, USA\\
$^{15}$  Department of Astronomy, University of Maryland, College Park, 
    MD 20742-2421, USA\\
$^{16}$  IRAM, Avenida Divina Pastora, 7, Núcleo Central, 
      E-18012 Granada, Spain\\
            }

\ead{eckart@ph1.uni-koeln.de}

\begin{abstract}
We report on recent near-infrared (NIR) and X-ray observations of
Sagittarius~A* (Sgr~A*), the electromagnetic manifestation of the
$\sim$4$\times$10$^6$\solm super-massive black hole (SMBH) at the Galactic
Center.  The goal of these coordinated multi-wavelength
observations is to investigate the variable emission from Sgr~A* in
order to obtain a better understanding of the underlying physical
processes in the accretion flow/outflow.
The observations have been
carried out using the NACO adaptive optics (AO) instrument at the
European Southern Observatory's Very Large Telescope (July 2005, May
2007) and the ACIS-I instrument aboard the \emph{Chandra X-ray
  Observatory} (July 2005).  We report on a polarized NIR flare
synchronous to a 8$\times$10$^{33}$~erg/s X-ray flare in July 2005,
and a further flare in May 2007 that shows the highest sub-flare to
flare contrast observed until now.  The observations can be
interpreted in the framework of a model involving a temporary disk
with a short jet.  In the disk component flux density variations can
be explained due to hot spots on relativistic orbits around the
central SMBH.  The variations of the
sub-structures of the May 2007 flare are interpreted as a variation of
the hot spot structure due to differential rotation within the disk.
\end{abstract}

\section{Introduction}

The investigation of the dynamics of stars has provided compelling
evidence for the existence of a super massive black hole (SMBH) at the
center of the Milky Way.  At a distance of only $\sim$8~kpc a SMBH of
mass $\sim$4$\times$10$^6$\solm can convincingly be identified with
the compact radio, infrared, and X-ray source Sagittarius A* (Sgr~A*;
Eckart \& Genzel 1996, Genzel et al. 1997, 2000, Ghez et al. 1998,
2000, 2004ab, 2005, Eckart et al. 2002 , Sch\"odel et al. 2002, 2003,
Eisenhauer et al. 2003, 2005). Additional strong evidence for a SMBH
at the position of Sgr~A* came from the observation of flare activity
on hourly time scales both in the X-ray and NIR wavelength domain
(Baganoff et al., 2001; Genzel et al., 2003; Ghez et al., 2004).

Due to its proximity Sgr~A* provides us with a unique opportunity to
understand the physics and possibly the evolution of SMBHs at the
nuclei of galaxies.  Sgr~A* is remarkably faint ($\leq10^{-9}$ of the
Eddington rate) in all wavebands. Its surprisingly low luminosity has
motivated many theoretical and observational efforts in the past
decade to explain the processes that are at work in the immediate
vicinity of Sgr A*.  By now, it is generally accepted that its feeble
emission is due to a combination of a low accretion rate with a low
radiation efficiency.  An intense discussion among the theoretical
community at present focuses on radiatively inefficient accretion flow
and jet models.  For a recent summary of accretion models and variable
accretion of stellar winds onto Sgr A* see Yuan (2006), Cuadra \&
Nayakshin (2006).  

The first successful simultaneous NIR/X-ray campaigns combined NACO
and Chandra as well as mostly quasi-simultaneous mm-data from BIMA,
SMA, and VLA (Eckart et al. 2004, 2006a).  The NIR/X-ray variability is
probably also linked to the variability at radio through
sub-millimeter wavelengths showing that variations occur on time
scales from hours to years (Bower et al. 2002, Herrnstein et al. 2004,
Zhao et al. 2003, 2004, Mauerhan 2005, Marrone et al. 2008,
Yusef-Zadeh et al. 2008).

The temporal correlation between rapid variability of the
near-infrared (NIR) and X-ray emission suggests that the emission
showing arises from a compact source within a few ten Schwarzschild
radii of the SMBH (Eckart et al. 2004, Eckart et al. 2006a).  In this
work,  we assume for Sgr~A* $R_s$=2$R_g$=2GM/c$^2$$\sim$8~$\mu$as, with $R_s$
being one Schwarzschild radius and $R_g$ the gravitational radius of
the SMBH.  For several simultaneous flare events the authors
found no time lag larger than an upper limit of $\le$10~minutes,
mainly constrained by the required binning width of the X-ray data.  The
flaring state can be explained with a synchrotron self-Compton (SSC)
model involving up-scattered sub-millimeter photons from a compact
source component.  Inverse Compton scattering of the THz-peaked flare
spectrum by the relativistic electrons then accounts for the X-ray
emission.  This model allows for NIR flux density contributions from
both the synchrotron and SSC mechanisms.  Observations for red and
variable NIR flare spectra (Eisenhauer et al. 2005, Hornstein et al. 2007,
Gillessen et al. 2006) are indicative of a possible exponential cutoff
of the NIR/MIR synchrotron spectrum (Eckart et al. 2004).

There is also evidence for a modulation of the NIR emission that may
be due to hot spots orbiting Sgr~A* in the accretion flow (Genzel et
al. 2003, Eckart et al. 2006b, 2008, Meyer et al.2006ab, 2007, Karas et
al. 2008).  The NIR flare emission is polarized with a well limited
range over which the position angle of the polarized emission is
changing (60$^o$$\pm$20$^o$) (Eckart et al. 2006b, Meyer et
al. 2006ab, 2007).  All these observations can be explained within a
model of a temporary accretion disk that occasionally contains
  one or several bright orbiting hot spot(s), possibly in conjunction
with a short jet, and suggest a stable orientation of the source
geometry over the past few years.

The millimeter/submillimeter wavelength polarization of Sgr A* 
is variable in both magnitude and position angle 
on timescales down to a few hours. 
Marrone et al. (2007)
present simultaneous observations made with the Submillimeter Array 
polarimeter at 230 and 350 GHz with sufficient sensitivity to 
determine the polarization and rotation measure at each band.
From their measurements they deduce an accretion rate that does not vary by
more than 25\% and - depending on the equipartition constraints and the 
magnetic field configuration -
amounts to 2$\times$10$^{-5}$ to 2$\times$10$^{-7}$ \solm yr$^{-1}$.
The mean intrinsic position angle is 167$^\circ$$\pm$7$^\circ$
with variations of $\sim$31$^\circ$ that must originate in the 
sub-millimeter photosphere of SgrA*.

Here, we present data and modeling for three events: polarimetric
  NIR observations of a very bright flare from May 2007, and
  \emph{Chandra} X-ray measurements from 2005 and 2004 that were taken
  in parallel with NIR photometric and polarimetric measurements of
  flares reported by Eckart et al.\ (2006ab).  In Section~2 we
summarize the observations and the data reduction.  The observational
results and modeling of the data are presented in Section~3 and a more
general discussion of available infrared and X-ray variability data on
Sgr~A* is given in Section~4.  In section~5 we briefy discuss the
  interaction of the GC ISM with a potential wind/partly collimated
  outflow that originates in the vicinity of Sgr~A*. In section~6 we
summarize our findings and draw some conclusions.

\section{Observations and Data Reduction}
\label{section:dataresults}

\begin{figure}[htb]
\centering
  \includegraphics[width=15cm,angle=00]{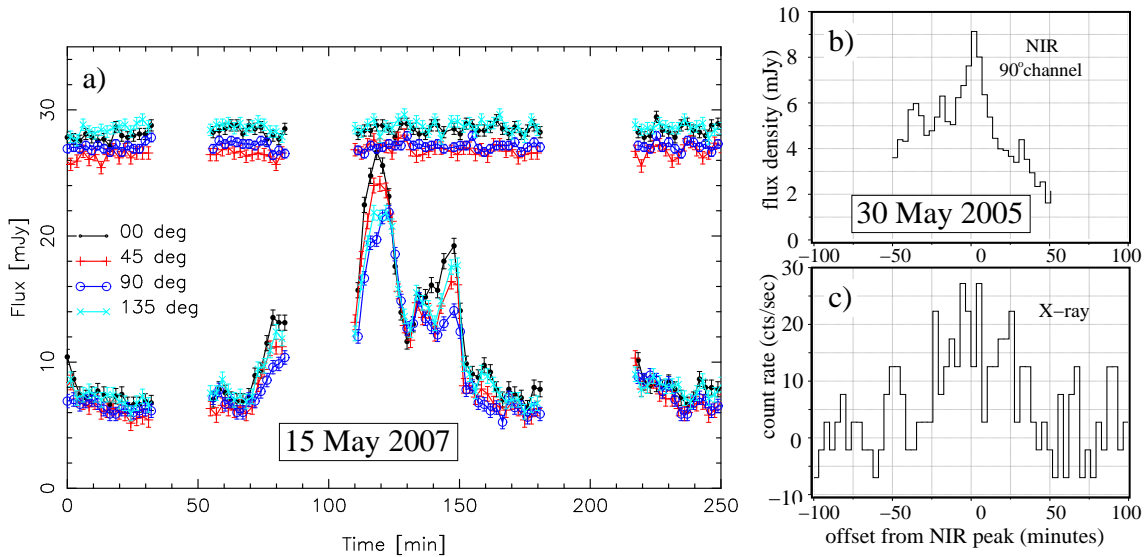}
\caption{{\bf a):} A plot of the flux density vs.\ time for of the
  exceptionally bright flare of SgrN A* observed with NACO/VLT on 15
  May 2007. Each polarimetri channel is depicted in a different color.
  The light curve of a constant star, S2, is shown in the same plot
  and shifted by a few mJy for clarity.  {\bf b) and c):}The NIR (top)
  and X-ray (bottom) data for the Sgr~A* flare observed on 30 May
  2005. To highlight both the flare and the sub-flare structure we
  plot only the flux in the NIR 90$^o$ polarization channel.  Both
  data sets have been sampled into bins of 207\,s width 
  The X-ray data are corrected for the intermediate quiescent emission.  
  The peak of the NIR fare occurred at 02:56:00 UT $\pm$3 minutes, 
  within about $\pm$7 minutes of the X-ray peak.  } \label{efig1}
\end{figure}

Embedded in coordinated, multi-wavelength campaigns, Sgr~A* was
observed in July 2005 and May 2007 , using the VLT\footnote{Based on
  observations at the Very Large Telescope (VLT) of the European
  Southern Observatory (ESO) on Paranal in Chile; Program:
  075.B-0093}.  Simultaneous X-ray observations were carried out with
the Chandra observatory in july 2005 (see below).  The NIR lightcurves
of Sgr~A* are depicted in Fig.~\ref{efig1}.  Details of
the observations and data reduction are given in Eckart et al. (2008)
and Eckart et al. (2006b).

 \begin{figure}
   \centering
   \includegraphics[width=8cm]{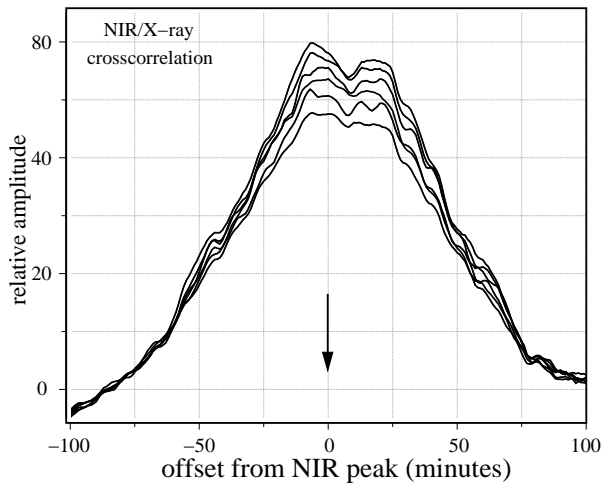}
      \caption{
Results of the cross correlation of the 207-second binned \emph{Chandra} X-ray data
with the polarized emission seen at position angles of (top to bottom)
90$^\circ$, 120$^\circ$, 60$^\circ$, 150$^\circ$, 30$^\circ$, 00$^\circ$.
The arrow marks the time offset of 0 minutes.
}
         \label{fig3}
   \end{figure}

The \emph{Chandra} X-ray data fully cover the observed polarized NIR
flare that we observed at the VLT in July 2005.  The X-ray data show a
8$\times$10$^{33}$~erg/s flare that is about 3 times as bright as the
quiescent emission from SgrA*.  In the right panels of
Fig.~\ref{efig1} we show corresponding X-ray and NIR lightcurves using
a 207 second bin size.
The cross-correlation of the X-ray data with the flux densities in the
individual NIR polarization channels shows that the flare event
observed in the two wavelength bands is simultaneous to within less
than 10 minutes.  The two sub-peaks in the cross-correlation function
correspond to two apparent sub-peaks in the X-ray light curve that
can, however, not be taken as significant given the SNR of $\sim$3
cts/s per integration bin 
In the X-ray domain there is no clear indication for a
sub-flare structure as observed in the NIR.  The NIR sub-flare
contrast defined as the sub-flare height divided by the height of the
overall underlying flare flux density ranges between 0.3 and 0.9.

\section{Modeling}
\label{ModelingResults}

\subsection{Basic building block of relativistic disk modeling of the flares}
\label{Relativistic-disk-modeling}
\label{sec:KY}

We interpret our polarized infrared flare events via the emission of
spots on relativistic orbits around the central SMBH in a temporary
disk (Eckart et al. 2006b, Meyer et al. 2006ab, 2007).  The model
calculations are based on the KY-code by Dovciak, Karas, \& Yaqoob
(2004) and are usually done for a single spot orbiting close to the
corresponding last stable orbit.  The possibility to explore effects
of strong gravity via time-resolved polarimetrical observations of
X-rays (which also inspired writing the KY code) was originally proposed by
Connors \& Stark (1977). The amplification light curves for
individual hot spots that can be computed with the KY code are used as
the basic building blocks of our models, 
because even a complicated (non-axisymetric) pattern on the
disk surface can be represented as a suitable combination of
emitting spots. At this point we just remind the reader that
relativistic effects actually do not produce polarization by
themselves, rather they can change the polarization angle and the
the overall polarization degree of an intrinsically polarized
signal because each photon experiences a different
gravitational effect along its path from the point of emission to
the observer. In case of a single spot as a source of the
emission, the observed polarization vector is expected to wobble
or rotate as a function of the spot phase. This is a purely
geometrical effect connected with the presence of strong
gravitational field. Naturally, the intrinsic changes of the spot
polarization are superposed on top of this relativistic effect.

\begin{figure}[t]
  \centering
  \includegraphics[width=10cm]{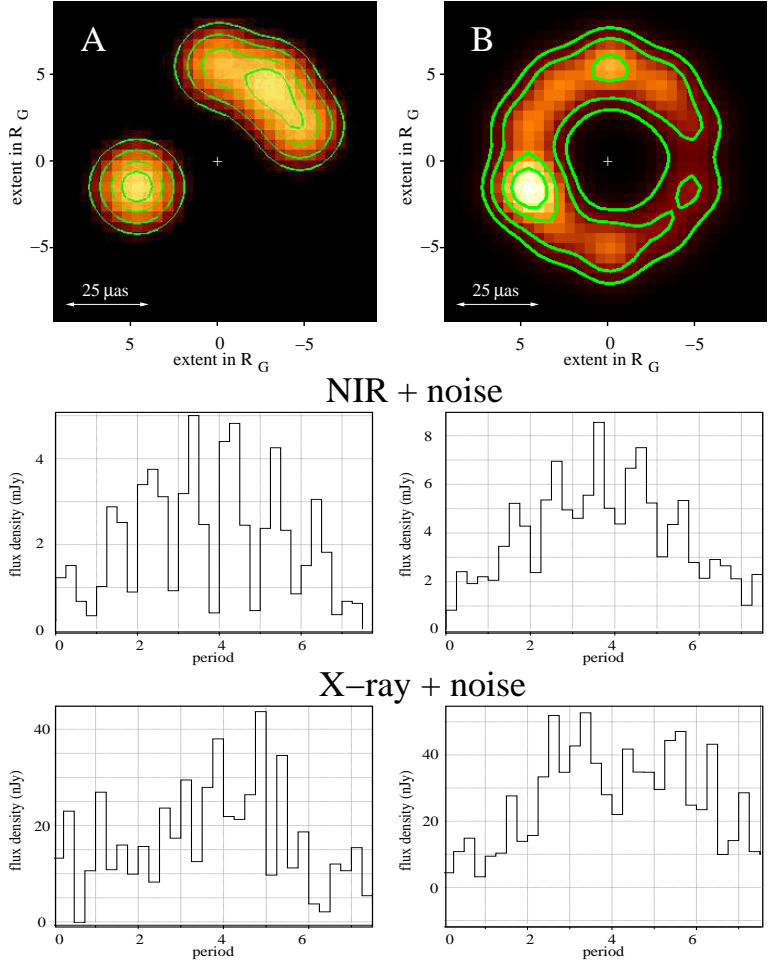}
     \caption{
As a result from model calculations, we show for two cases
representative flux density distributions and NIR/X-ray light model
curves with noise.  
Different distribution of spots in the disk were assumed in both cases.
The flux density distributions are shown along the
last stable orbit perimeter of the super massive black hole associated
with Sgr~A* (upper panel).  Here, no truncation at or just within the
last stable orbit has been applied.
The contour lines are at 12, 25, 50, and 75\% of the peak of 
the flux density distribution.  The NIR
and X-ray light curves shown in the lower panels are representative
for the median values calculated in Eckart et al. (2008).  
For the X-ray data we added noise comparable to the noise contributions
obtained in the observations.
For the NIR we added 0.4\,mJy of random Gaussian noise. The
bin size of the model data corresponds to 207\,s for an assumed
14\,min period. The position of Sgr~A* (not visible in the figure)
is indicated by a white cross.
  }
        \label{efig21}
  \end{figure}

\begin{figure}[t]
  \centering
  \includegraphics[width=10cm]{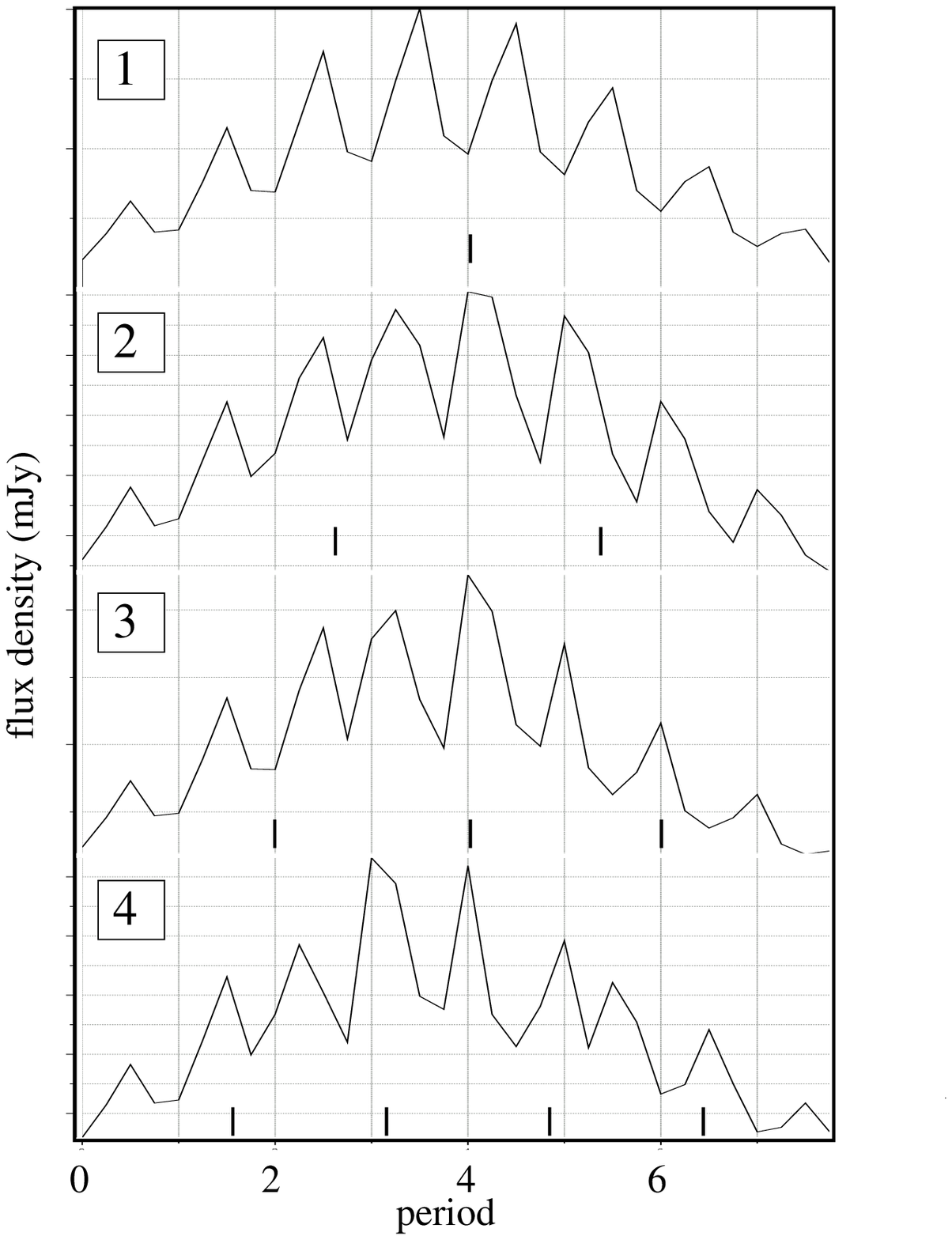}
     \caption{
Simulation of 2.2$\mu$m light curves under the assumption
of decreasing times scales for the stability of source components in
the accretion disk around SgrA*.  The labels and thin vertical lines
mark the number and centers of Gaussian shaped stability time
intervals.  These marks are spaced by a FWHM of the individual
distributions.  For short stability times scales the overall
appearance of the light curve is preserved but the sub-flare
amplitudes and time separations vary.
  }
        \label{efig22}
  \end{figure}

\begin{figure*}[t]
  \centering
  \includegraphics[width=15.0cm]{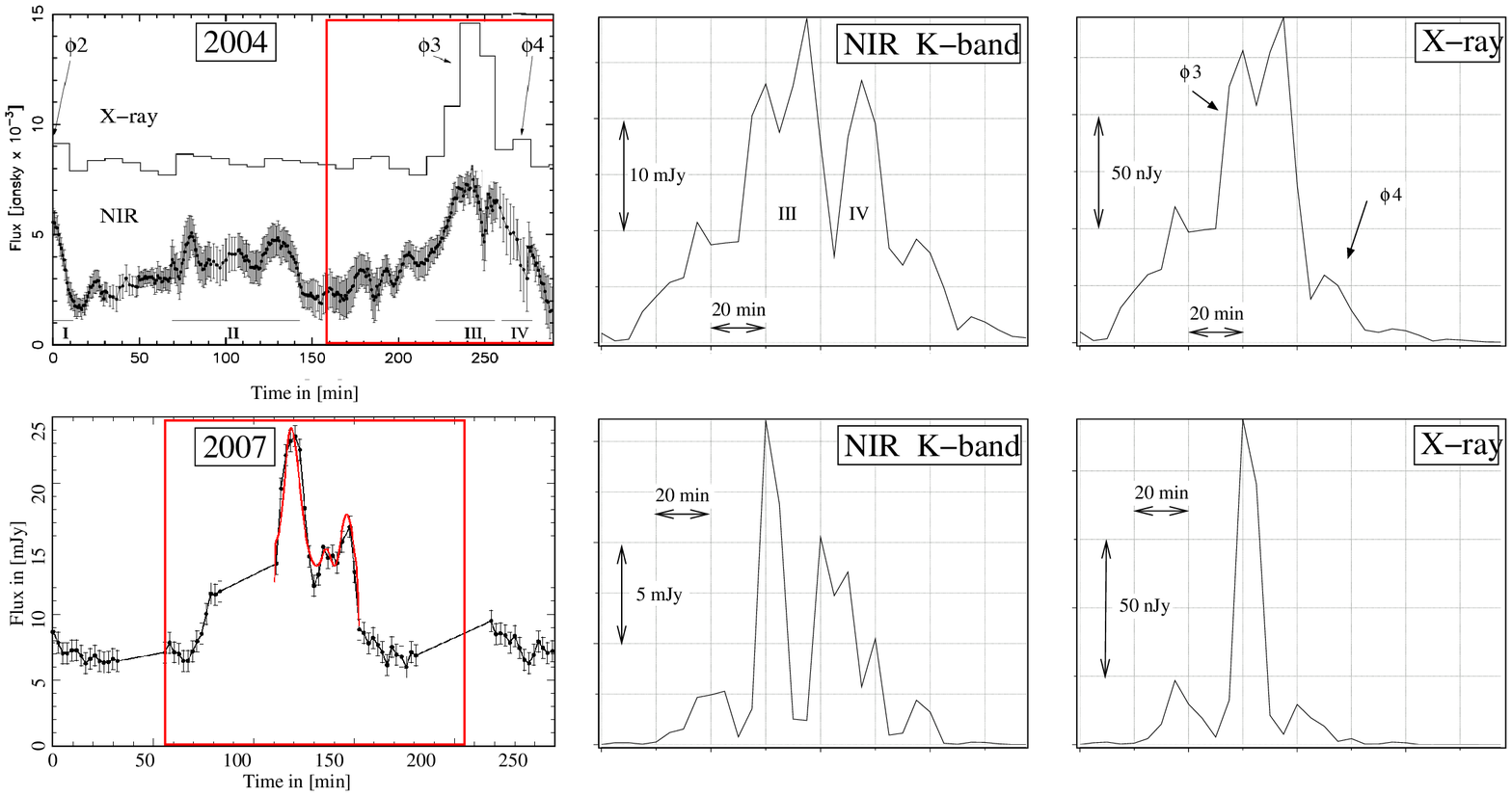}
     \caption{
Application of the time dependent flare emission model presented in
Section \ref{section:diskmodel} to the data obtained in 
May 2007 (bottom) and July 2004 (top).
In the panels on the left we show the available NIR and X-ray data with
the modeled portion indicated by a red line. In the middle and on the
right we show the NIR K-band and X-ray light curve derived from a 
multi component disk model. In both cases, starting at the center 
of the flare event, we assumed a 30\% increase of the source component 
sizes over about 40 minutes i.e. two orbital time scales.
With the additional assumption of a flux decrease of $\sim$1 magnitude 
due to synchrotron losses the model provides a very good qualitative and
quantitative representation of the 2004 measurements (see Eckart et al. 2006a).
For the 2007 NIR data there are no simultaneous X-ray data available 
but the K-band light curve indicates an evolution of the source components.
The lower right panel therefore represents the light curve
we would have expected.
}
\label{efig3}
\end{figure*}

\subsection{A multi component disk model}
\label{multicomp}

The observed NIR/X-ray properties of the SgrA* light curves raise a
number of questions: Can we expect a sub-flare structure in the X-ray
domain using a synchrotron self-Compton model?  What is the
approximate flux distribution within a temporary accretion disk around
Sgr~A*?  This is also closely related to more general questions of how
the observed light curve properties vary if the life time of the spot,
shearing, and synchrotron cooling time scales are considered.  In the
following we describe an extended SSC model that includes a disk
structure that is composed of a combination of hot spots of
  different brightness and with different initial orbital locations
  (see section above).  Our model allow us to calculate light curves
in the NIR and X-ray domains in order to discuss the questions posed
above.  A detailed description of the used SSC model is given in
Eckart et al. (2006a).

\subsection{The SSC disk model}
\label{section:diskmodel}

In order to explain the time dependent flare properties we 
assume that the sub-flare and disk component can be described by
a number of individual synchrotron and SSC emitting source components.
Combining the light amplification curve for individual orbiting spots
and a simple SSC model, we can obtain zero order 
time dependent flare characteristics from the NIR to the 
X-ray domain.

As a starting point we used synchrotron models that represent a high
flux density, i.e. flaring, and a low flux density state.  Greenhough
et al. (2001) outline the importance of scaling properties of the
transport processes operating within accretion disks.  Pessah et
al. (2007) present a scaling law between magnetic stress in units of
the gas pressure and the vertical disk cell size in units of the
pressure scale height implying that the magnetic field and source
component size follow a power law relation.  Therefore we assume that
the essential quantities of the SSC models, i.e. the turnover flux
density $S_m$ and frequency $\nu_m$ as well as the source size
$\theta$ of the individual source components are distributed as power
laws with the boundary values taken from the high and low flux density
state models. The corresponding power-law indices of the distribution are
$\alpha_{S}$, $\alpha_{\nu}$ and $\alpha_{\theta}$:

$$N(S_{m}) \propto S_m^{\alpha_S}~~~,~~~
N(\nu_{m}) \propto \nu_m^{\alpha_\nu}~~~,~~~
N(\theta) \propto \theta^{\alpha_\theta}~~.$$

For example, if $\alpha_S=0$ the flux densities of the source
components are distribted over the full range between the minimum
  and maximum values with equal frequencies. For $\alpha_S>0$ and
$\alpha_S<0$ there is an increasing preference towards larger and
lower flux density values, respectively.  Similarly this is true for
$\alpha_{\nu}$ and $\alpha_{\theta}$.

The innermost stable circular orbit (ISCO) around a non-rotating black
hole with spin parameter $a$=0 is 6$R_g$.
Assuming the co-rotating case, that radius will shrink for higher spin
parameters.  For a rotating black hole with $a$=0.5 the ISCO is
$\sim$4.4$R_g$.  Model calculations have shown (Meyer et al. 2006ab)
that for Sgr~A* spin parameters $a\ge0.5$ and source components
orbiting at radii larger than the ISCO are very likely 
It was furthermore shown that the disk is
small with an outer disk radius extending not much further than 2$R_s$
beyond the ISCO.  With source component sizes of the order of
1.5~$R_g$ (Meyer et al. 2006ab) we can safely assume that the disk is
well sampled using a total of 10 Gaussian shaped disk sections with
random values of $S_m$, $\nu_m$ taken from the described
power law distributions in order to model the entire accretion disk
(see Eckart et al. 2008 for more details).
The brightest
of these sections will then represent the orbiting spot and the rest
will account for the underlying disk.  This setup will, of course,
also allow for several bright spots.  As a simple - but still general
- model we assumed the source components to be equally spaced along
the circumference of a constant orbit.  While orbiting, the flux
density of each component will follow the achromatic magnification
curves that can be calculated as a function of the spin parameter $a$,
inclination $i$ and orbital radius with the KY code as described
 in section\,\ref{sec:KY}.  In  order to model the limited
  lifetims of the spots we  apply a Gaussian
shaped weighting function with a FWHM of about 3 orbital periods,
which resembles the observed flare lengths quite well.
The combination of these weighting functions cover the overall flare.
The model implies the
that the life time of spots is short with respect to the orbiting time scale 
and the overall flare duration and that new hot spots have to be 
created.
It is not yet clear what
kind of process would be the best one to describe the creation
and subsequent extinction of the spots. The assumption that spots
arise as being statistically independent on each other is a reasonable
first approximation, however, it is quite likely that some kind
of relationship between the spots will have to be included,
for example within the framework of the avalanche mechanism
(Pechacek et al. 2008).

\subsection{Results of the Modeling}
\label{section:modelresults}

An important result of the simulations is that not only the
observed total NIR and X-ray flux densities can be successfully
modeled  but also the observed sub-flare contrast.  In addition,
the best fits to the NIR and X-ray flux densities lie within or close
to regions of high NIR flux density weighted magnetic field strength.
Under the assumption that the NIR polarization measurements are being used
- the NIR flux density weighted magnetic 
field strength then represents the magnetic field value that is most
likely  to be measured.
This demonstrates that
the combination of the SSC modeling and the idea of a temporary
accretion disk can realistically describe the observed NIR polarized
flares that occur synchronous with the 2-8~keV X-ray flares.  

We find that the power-law index $\alpha_S$ of the assumed power law
distribution for the synchrotron peak flux S$_m$ results in best model
results for values around $\alpha_S$=-1$\pm$1.  A value of
$\alpha_S$=0 (which is included in this range) represents
secnarios in which source components cover the
entire range of possible flux densities with an equal probability for
each value rather than being biased in a way that the components
  have a large probability of having similar brightness.  Values
  of $\alpha_S\approx$0 provide high sub-flare contrast values.  An
exponent of $\alpha_S$=-1 favors a higher frequency of lower flux
density values.  In the SSC model high contrast is provided by the SSC
contribution to the NIR spectral range, also allowing for
$\chi^2$-fits at lower flux density weighted magnetic field strengths
around 30~G rather than 60~G as for the synchrotron model.

Magnetic field strengths between
5 and 70 Gauss (Eckart et al. 2006a, 2008, Yusef-Zadeh et al. 2008) are consistent with
sub-mm/mm  variability timescales of synchrotron components with
THz peaked spectra and the assumption that these source components 
have an upper frequency cutoff $\nu_2$ in the NIR, i.e. that they contribute significantly to
the observed NIR flare flux density.
Here the upper frequency cutoff to the synchrotron spectrum is assumed to be at 
$\nu_2 = 2.8 \times 10^6 B \gamma_2$ in Hertz, with the magnetic field strength in Gauss.
The Lorentz factor $\gamma_2$ corresponds to the energy $\gamma_2 mc^2$ at the upper
edge of the electron power spectrum.
For $\gamma_2$$\sim$10$^3$ and B around $60~G$, the synchrotron cutoff falls into the NIR.
In order to match the overall typical flare timescale of about
2 hours and given a minimum turnover frequency around 300~GHz 
the minimum required magnetic field strength 
is of the order of a 5 Gauss.
This is required as a minimum value to have the cooling time of the overall flare less than the 
duration of the flare (Yuan, Quataert, Narayan 2003, 2004, Quataert 2003).

After adding appropriate noise (estimated from the available
X-ray data) to the X-ray modeling results it becomes apparent that
at the given SNR and data sampling short term variability in the
X-ray data are difficult to determine, even if they have a
modulation contrast similar to that observed in the NIR (see
Fig.~\ref{efig21}).
Bright spots may on average have smaller sizes or lower cutoff
frequencies.  An increase of SSC X-ray flux density due to an increase
of THz peak synchrotron flux may be compensated by this effect.  Hence
the sub-flare contrast may be much lower in the X-ray compared to the
NIR domain.

In Fig.~\ref{efig3} we show the modeling results for the May 2007 NIR
and the July 2004 (see Eckart et al., 2006a) simultaneous NIR/X-ray
data on SgrA* using our time dependent flare emission model.  For the
2007 data we implemented a double hot spot model in the KY code that
was briefly described in Section~\ref{Relativistic-disk-modeling}.
For the 2004 data we involved a model consisting of 7 components at
increasing distances from the SMBH starting at the inner last stable
orbit.  The components line up close to the line of sight 
on opposite sides of the SMBH close to the flare center in time. 
This corresponds to a relative minimum in the light curve.
This arrangement of source components therefore provides a maximum 
amount of Doppler amplification before and after this alignment 
by either one or the other component.
This gives rise to the two NIR flare events labeled III
and IV (Eckart et al. 2006a).  

Motivated by the fact that the May 2007 data show evidence for hot
spot evolution due to differential rotation within the relativistic
disk, we assumed that an increase of the source size of the
  individual spots may be of importance during the flare events.
Therefore, starting at the center of the July 2004 flare event,
we assumed in both cases a 30\% increase of the source component sizes
over 30 to 40 minutes, i.e. about two orbital time scales.  This
results in a sharp decrease of the SSC X-ray flux density and
therefore in a very good representation of the 2004 measurements (see
X-ray flares labeled $\phi$3 and $\phi$4 in Fig.~\ref{efig3} and
Eckart et al. (2006a).  Based on these time dependent model
assumptions we would have expected a similar evolution of the
X-ray flare light curve for the May 2007 NIR observations as shown in
the bottom right panel of Fig.~\ref{efig3}.

Such a scenario may also explain the 2006 July 17 Keck NIR/X-ray light
curves reported by Hornstein et al. (2007).  The authors measured an
NIR flare without a detectable X-ray counterpart.  It was delayed by
about 45 minutes from a significant X-ray flare, during which no NIR
data were taken.  Assuming that the X-ray flare was accompanied by an
unobserved NIR flare as well, this event may have been very similar in
structure to the July 2004 flare.

\subsection{\bf Sub-flares and quasi-periodicity}

For Fig.~\ref{efig22} we calculated 2.2$\mu$m light curves
under the assumption of decreasing times scales for the
magneto-hydro-dynamical stability of the source components in the
accretion disk around SgrA*.  The thin vertical lines mark the centers
of stability intervals with Gaussian shaped flux density weights.
These Gaussian shaped weights cover a time time interval over which 
source components can be considered as being stable.
Between such time intervals new source components within the 
accretion disk are being formed.
The marks are spaced by a FWHM of the individual Gaussians.
This arrangement results in light curves similar to
the observed ones (see Eckart et al. 2008 for more details).  
We assumed that for
each of these intervals the flux density distribution within the disk
is different. This results in phase shifts between the light curves
(i.e. different positions of the spot within the disk) of $\pm$$\pi$.
This simulation shows that the overall appearance (especially the mean
QPO frequency) of the light curve can be preserved and that variations
in the sub-flare amplitude and time separations can be explained by
such a scenario.  In Fig.~\ref{efig22} both quantities
vary by a factor of 2. Larger variations are possible for stronger
variations of the spot brightness.

In a simple model by Schnittman (2005) hot spots are created and
destroyed around a single radius with random phases and exponentially
distributed lifetimes $T_{lif}$, resulting in Lorentzian peaks in the
power spectrum at the orbital frequency with a width of
$\Delta$$\nu$=(4$\pi$$T_{lif}$)$^{-1}$.  The typical lifetimes of
spots in this model are proportional to the orbital period $T_{orb}$
at their radius.  From MHD calculations Schnittman et al. (2006) find
over a large range of radii that disk perturbations indeed have a
nearly exponential distribution of lifetimes, with $T_{lif}$$\sim$
0.3$T_{orb}$.  This implies that even if the spot lifetime is solely
determined by the cooling time at 2.2 or 3.8~$\mu$m, this scenario is
in full agreement with the suggested quasi-periodicity since
T$_{lif}$$\sim$0.3T$_{orb}$$\sim$$T_S$.
Here  $T_S$ is the synchrotron cooling time in the NIR 
which is of the order of 
several minutes for the magnetic filed strengths used here
(see Eckart et al 2008).

Assuming that the width of the observed QPO is 17$\pm$3~minutes we can
derive an expected full width of the power spectrum peak of
$\Delta$$\nu$$\sim$0.02~min$^{-1}$.  Following Schnittman et
al. (2005) this corresponds to an expected lifetime of the spots of
$T_{lif}$$\sim$4~minutes - a value similar to the synchrotron cooling
time in the NIR K-band.  However, quasi-simultaneous K- and L-band
measurements by Hornstein et al. (2007) show that for several 1 to 2
hour stretches of variable K-band emission $\ge$3~mJy, including
flares of 10 to 30 minutes duration, the light curves at both
wavelengths are well correlated.  This suggests that the synchrotron
cooling time scale in this case appears to be longer than the
  flare time scale and therefore not to be a relevant quantity for the
spot lifetime.  In addition the spread $\Delta$$\nu$ is an upper
limit to the width of a possible Lorentzian distribution describing
the QPO measurements.  Therefore, we have to assume that $T_{lif}$ is
even longer than the synchrotron cooling times at K- and L-band, i.e
significantly longer than 13~minutes, and suggesting that the spot
lifetime could be of the order of $T_{orb}$, in agreement with results
by Schnittman et al. (2005) and the model calculations presented here.
The synchrotron cooling time scales may not be relevant at K- and
L-band if the heating time scale is longer (e.g. on the time scale of
the overall NIR or sub-mm flare event) or if some additional mechanism
is at work that stabilizes the spots in the temporary accretion disk
of SgrA*.  A small spot size and a high magnetic field intrinsic to
the spot may help to prevent strong shearing, lowering the
requirements on this confinement mechanism.

\section{Alternative models to the black hole scenario}

Explaining SgrA* with alternative solutions for a MBH becomes
increasingly difficult (see discussion in appendix of Eckart et
al. 2006b).  Stellar orbits near SgrA* make a universal Fermion ball
solution for compact galactic nuclei highly unlikely and especially
the fact that SgrA* appears to be a strongly variable and mass
accreting object, represents a problem for the stability constraints
that boson or fermion balls have.  It is, for instance, quite a
delicate process to form a boson star and preventing it from
collapsing to a super massive black hole despite of further accretion
of matter, a non spherically symmetric arrangement of forces as in the
case of a jet or matter being in orbit around the center but well
within the boson star.  Such a massive boson star scenario could
already be excluded for the nucleus of MCG-6-30-15 (Lu \& Torres
2003).  In the case of a stationary boson star the orbital velocity
close to the $\sim$3~R$_S$ radius LSO is already $\sim$3 times lower
than that of a Schwarzschild MBH (Lu \& Torres 2003) and relativistic
effects are severely diminished and further reduced at even smaller
radii.  If the variability that is indicated especially by
the infrared polarization data is indeed due to orbital motion of a spot
within a temporal accretion disk then a stationary boson star can be
excluded as an alternative solution for SgrA*, since in this case one
expects the orbital periods to be larger.

\section{Interactions of the GC ISM with a central wind}

L'-band (3.8 $\mu$m) images of the Galactic Center show a large number
of thin filaments in the mini-spiral, located west of the mini-cavity
and along the inner edge of the Northern Arm (Muzic et al. 2007).
Only a few of these filaments are associated with stars.  Similar
filaments have been seen in high-resolution radio data (Zhao \& Goss
1998) and NIR Pa~$\alpha$ (Scoville et al. 2003), Br~$\gamma$ (Morris
2000) and HeI (Paumard et al. 2001) emission line maps.  One possible
mechanism that could produce such structures is the interaction of a
central wind with the mini-spiral.  Muzic et al. (2007) present the
first proper motion measurements of the thin filaments observed in the
central parsec around SgrA* and investigate possible mechanisms that
could be responsible for the observed motions.  The observations have
been carried out using the NACO adaptive optics system at the ESO VLT.

\begin{figure}
\centering
\includegraphics[width=15cm]{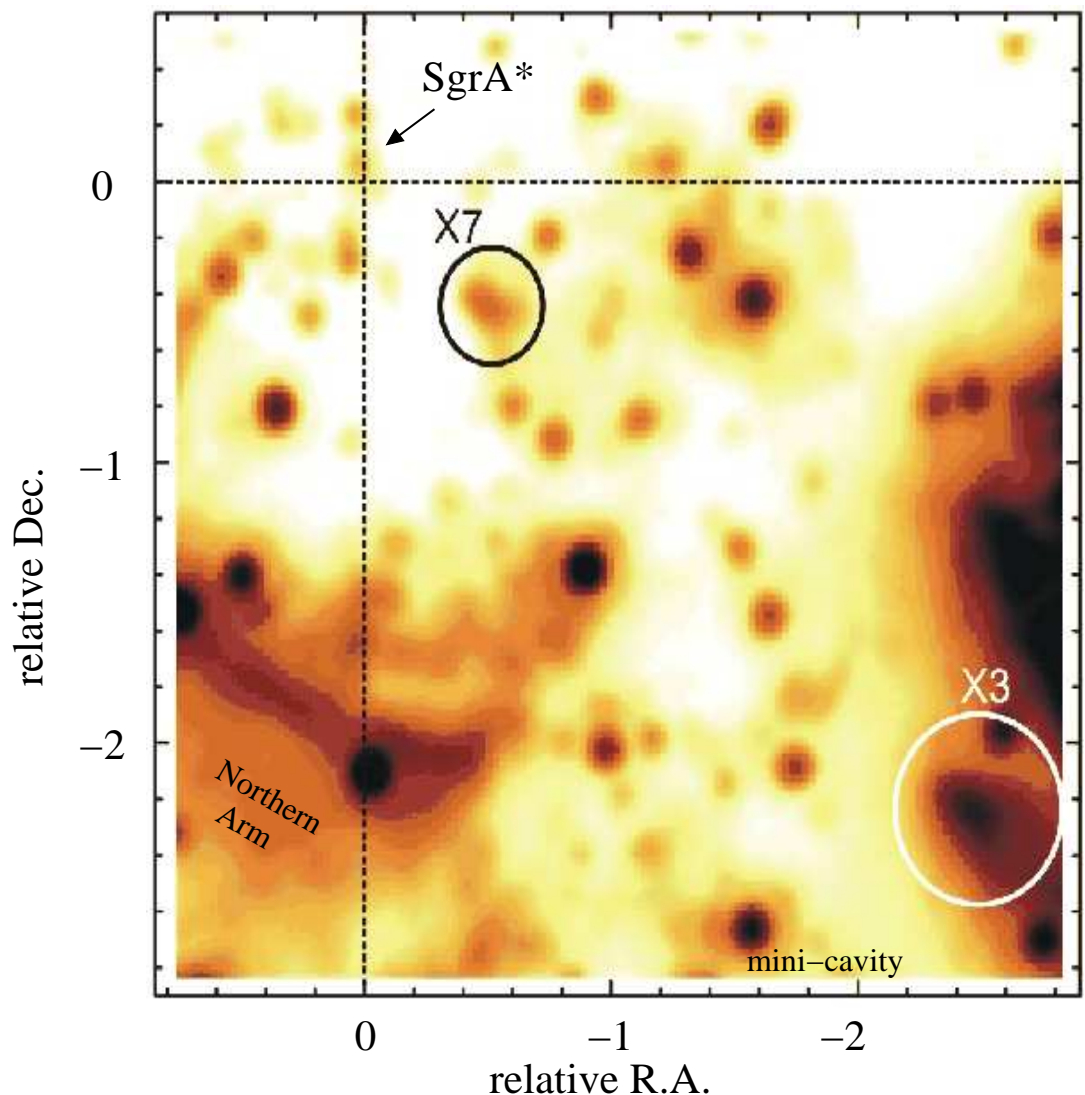}
\caption{
A   3.8``$\times$3.6``  section of an L'-band image including
SgrA* and the two cometary shaped sources X3 and X7.
The V-shaped dust shells indicate an interaction with a strong wind
in the local Galactic Center ISM. 
The V-shapes of both sources are pointed toward the position of SgrA*. 
This suggests that the wind originates in the immediate surrounding of SgrA*.
}
\label{efig4}
\end{figure}

They show that the shape and the motion of the filaments does not
agree with a purely Keplerian motion of the gas in the potential of
the supermassive black hole at the position of SgrA*.  Therefore,
additional mechanisms must be responsible for their formation and
motion.  The authors argue that the properties of the filaments are
probably related to an outflow from the disk of young mass-losing
stars around SgrA*.  In part, the outflow may originate from the black
hole itself.  They also present some evidence and theoretical
considerations that the outflow may be collimated.

Muzic et al. (2007) also derive the proper motions of 2 cometary shaped dusty
sources close (in projection) to SgrA* (Fig.\ref{efig4}).  The
V-shaped dust shells indicate an interaction with a strong wind in the
Galactic Center ISM (Fig.\,\ref{efig4}).

The central cluster of massive stars provides $\sim~$3$\times$10$^{-3}$\solm
yr$^{-1}$ of gas in the form of stellar winds to the center (Najarro
et al. 1997).  However, about 99\% of the material from stellar winds
does not even get close to the Bondi radius and must therefore escape
the central arcseconds in form of a wind and only a small fraction of
the gas is actually accreted onto the black hole (Baganoff et
al. 2003, Bower et al. 2003, Quataert 2003, Marrone et al. 2006; see
also ADIOS and RIAF theory e.g. Blandford \& Begelman 1999 and Yuan et
al. 2006).  The outflow may be partially collimated in form of a jet.
Jet and shock models (see Markoff \& Falcke 2003, Melia \& Falcke
2001) as well as accretion from an in-falling wind due to the
surrounding cluster of mass-losing hot stars (e.g. Coker et al. 1999,
Yuan et al. 2002) are discussed to explain the radio to X-ray
properties of SgrA*.


The V-shapes of both sources 
are pointed toward the position of SgrA* and therefore represent the
most direct indication for a wind from SgrA*.

\section{Summary and Conclusion}

We have detected an X-ray flare that occurred synchronous to a NIR
flare with polarized sub-flares. This confirms the previous finding
(Eckart et al. 2004, Eckart et al. 2006a, Yusef-Zadeh et al. 2006)
that there exists a class of X-ray flares that show simultaneous NIR
emission with time lags of less than 10~minutes.  In addition there
are lower energy flare events that are bright in the infrared and are
not detected in the X-ray domain (Hornstein et al. 2007).  In the
framework of a relativistic disk model the May 2007 polarimetric NIR
measurements of a flare event with the highest sub-fare to flare
contrast observed until now, may provide direct evidence for a
spot evolution during the flare.  This supports the interpretation of
the NIR polarimetry data within a relativistic disk model.  Combined
with the assumption of spot expansion due to differential rotation,
the combined SSC disk model can explain the  combined X-raw and
  NIR data of the July 2004 flare (Eckart 2006a) and possibly also of
the flare from 17 July 2006 reported by Hornstein et al. (2007).

\begin{figure}[t]
\centering
  \includegraphics[width=15cm,angle=00]{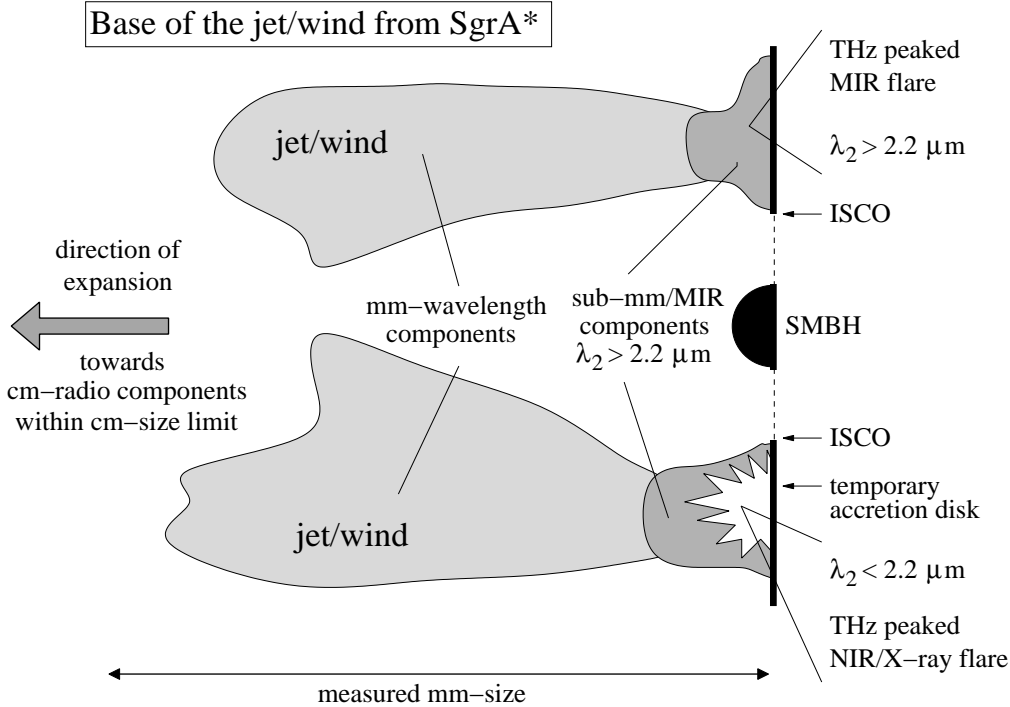}
\caption{A possible source structure for the accretion 
disk around the SMBH associated with Sgr~A*.
In this sketch the disk is shown as a vertical thick line to the right.
Extending to the left, we show one side above the disk.
Higher energy flare emission (lower part) is responsible for the observed
NIR/X-ray flare emission.
Lower energy flare emission (upper part) may be peaked in the THz domain
and may substantially contribute to long wavelength
infrared emission. Here we have assumed that their NIR/X-ray contributions are
negligible (see more details and model discription in Eckart et al. 2008).
In addition to the expansion towards and beyond the the mm-source size, r
adial and azimuthal
expansion within the disk may occur.
Here $\lambda_2$ is the wavelength corresponding to the upper synchrotron cutoff frequency $\nu_2$.
} \label{efig5}
\end{figure}

The combination of relativistic amplification curves with a simple SSC
mechanism allows us zero order interpretations in a time dependent
flare emission model.  We find that the temporary accretion disk
around Sgr~A* can well be represented by a multi component model with
source properties that are bracketed by those of a simple flare and a
quiescent model.  We have used a ($\gamma_e \sim 10^3)$ synchrotron
model in which the source component spectral indices are compatible
with the constant value of $\alpha=0.6\pm0.2$ reported by (Hornstein
et al. 2007).  A steeper spectral index of $\alpha=1.3$ allows for
direct synchrotron and SSC contributions in the NIR.  In both the
  July 2005 and the May 2007 flare the component flux densities can
be represented by a power spectrum $N(S) \propto S_m^{\alpha_S}$ with
an exponent $\alpha_S$ close to -1.  The multicomponent model explains
the sub-flare structure at infrared wavelengths and shows that with
adequate sensitivity and time resolution the sub-flares should be
detectable in the X-ray domain as well.

Eckart et al. (2008) present a model in which a combination of a
temporary accretion disk occurs in combination with a short jet.  Such
a source structure may explain most of the observed properties of
Sgr~A*.  Such a configuration is sketched in Fig.\ref{efig5}.  In this
figure the disk is seen edge-on.  Details of expected jet geometries
are discussed by Markoff, Bower \& Falcke (2007).

Simultaneous NIR K- and L-band measurements in combination with X-ray
observations should lead to a set of light curves that can allow us to
prove the proposed model and to discriminate between the individual
higher and lower energy flare events.  Simultaneous X-ray measurements
are important to clearly distinguish between high and low energy
events.  To do so, it is required to separate the thermal non-variable
bremsstrahlung and the non-thermal variable part of the Sgr~A* X-ray
flux density. This can only be achieved with a sufficiently high
  angular resolution in the X-ray regime.  This capability is
provided by the ACIS-I instrument aboard the \emph{Chandra X-ray
  Observatory} and is essential for the proposed observations,
especially in the case of weak X-ray flare events in which the X-ray
flare intensity is of the order of the extended bremsstrahlung
component associated with SgrA* - or even below.  These can clearly be
identified in combination with infrared data.

\section*{Acknowledgements}
Part of this work was supported by the German
\emph{Deut\-sche For\-schungs\-ge\-mein\-schaft, DFG\/} via grant SFB 494.
L. Meyer, K. Muzic, M. Zamaninasab, D. Kunneriath, and R.-S. Lu,
 are members of the International Max Planck Research School (IMPRS) for
Astronomy and Astrophysics at the MPIfR and the Universities of
Bonn and Cologne. RS acknowledges support by the Ram\'on y Cajal
programme by the Ministerio de Ciencia y Innovaci\'on of the
government of Spain.

\end{document}